\documentclass[12pt]{article}
\usepackage{a4wide,amssymb,cite,rotate,epsf}

\newcommand{\gf}{\gamma_5}
\newcommand{\wh}{\widehat}
\newcommand{\nn}{\nonumber}
\newcommand{\IM}{\mbox{\rm Im}}
\newcommand{\eqn}[1]{(\ref{#1})}
\newcommand{\mev}{\mbox{\rm MeV}}
\newcommand{\gev}{\mbox{\rm GeV}}
\newcommand{\FF}{\langle aFF\rangle}

\newcommand{\smvs}{\vbox{\vskip 8mm}}

\newcommand{\qq}{\langle\bar qq\rangle}
\newcommand{\MSb}{{\overline{{\rm MS}}}}

\newcommand{\newsection}[1]{\section{#1}\setcounter{equation}{0}}

\begin{document}


\begin{titlepage}
\begin{flushright}
{HD-THEP-01-34}\\[16mm]
\end{flushright}

\begin{center}
{\Huge\sf\bf\boldmath $f_B$ and $f_{B_s}$ from QCD sum rules
 \unboldmath}\\[12mm]

{\large\bf Matthias Jamin${}^{1,*}$}
{\large and}
{\large\bf Bj\"orn O. Lange${}^{2}$}\\[10mm]

{\small\sl ${}^{1}$ Institut f\"ur Theoretische Physik, Universit\"at
           Heidelberg,} \\
{\small\sl Philosophenweg 16, D-69120 Heidelberg, Germany}\\
{\small\sl E-Mail: M.Jamin@ThPhys.Uni-Heidelberg.DE}\\
{\small\sl ${}^{2}$ Physics Department, Clark Hall, Cornell University,} \\ 
{\small\sl Ithaca, NY 14853, USA}\\
{\small\sl E-Mail: bol1@cornell.edu}\\[12mm]
\end{center}

{\bf Abstract:}
The decay constants of the pseudoscalar mesons $B$ and $B_s$ are evaluated
from QCD sum rules for the pseudoscalar two-point function. Recently calculated
perturbative three-loop QCD corrections are incorporated into the sum rule. An
analysis in terms of the bottom quark pole mass turns out to be unreliable
due to large higher order radiative corrections. On the contrary, in the $\MSb$
scheme the higher order corrections are under good theoretical control and a
reliable determination of $f_B$ and $f_{B_s}$ becomes feasible.

Including variations of all input parameters within reasonable ranges,
our final results for the pseudoscalar meson decay constants are
$f_B = 210 \pm 19\,\mev$ and $f_{B_s} = 244 \pm 21\,\mev$.
Employing additional information on the product $\sqrt{B_B}f_B$ from global
fits to the unitarity triangle, we are in a position to also extract the
$B$-meson $B$-parameter $B_B = 1.26 \pm 0.45$. Our results are well compatible
with analogous determinations of the above quantities in lattice QCD.

\vfill



\vspace{4mm}
{\small ${}^{*}$ Heisenberg fellow.}
\end{titlepage}

\newpage
\setcounter{page}{1}


\newsection{Introduction}

Experimental efforts in recent years have provided us with a wealth of new
information on decays of bottom hadrons. To achieve a good understanding of
these data, also the impact of the strong interactions has to be controlled
quantitatively. This requires the accurate calculation of hadronic matrix
elements involving $B$-hadrons. Generally, hadronic matrix elements contain
contributions from low energies and thus non-perturbative methods should be
employed for their evaluation. Current approaches include lattice QCD, QCD
sum rules and the effective theory of heavy quarks (HQET). In this work, we
shall consider a calculation of the simplest type of hadronic matrix elements,
namely the pseudoscalar $B$- and $B_s$-meson decay constants $f_B$ and
$f_{B_s}$ in the framework of QCD sum rules \cite{svz:79,rry:85,nar:89,shi:92}.

The pseudoscalar decay constants parametrise $B$-meson matrix elements of the
axial\-vector current with the corresponding quantum numbers and are defined
by
\begin{equation}
\label{fBfBs}
\langle 0|(\bar q\gamma_\mu\gf b)(0)|B(p)\rangle \; = \; i f_B p_\mu
\quad \mbox{and} \quad
\langle 0|(\bar s\gamma_\mu\gf b)(0)|B_s(p)\rangle \; = \; i f_{B_s} p_\mu \,.
\end{equation}
Throughout this work we assume isospin symmetry and $q$ can denote an up or
down quark. Weak interactions induce the leptonic decay of the $B$-meson. For
example $f_B$ then appears in the decay width of the process $b\bar u \to
l\bar\nu_l$ which takes the form
\begin{equation}
\label{Gammal}
\Gamma(B^-\to l^-\bar\nu_l) \; = \; \frac{G_F^2}{8\pi}\,|V_{ub}|^2 f_B^2 \,
m_l^2 m_B \Biggl( 1-\frac{m_l^2}{m_B^2} \Biggr) \,,
\end{equation}
completely analogous to the corresponding decay of the light pseudoscalar
mesons. Despite the suppression by the small factors $m_l^2$ and $|V_{ub}|^2$,
there is some hope that the leptonic decay $B\to l\bar\nu_l$ can be measured
at the B-factories within the next years. Once $f_B$ is assumed to be known,
this would provide a very clean determination of $|V_{ub}|$. In any case, $f_B$
is an important quantity for it also enters more complicated hadronic matrix
elements of $B$-mesons like form factors or matrix elements of four-quark
operators.

The calculation of heavy meson decay constants in QCD has a rather long
history. For charmed mesons, they were first considered in \cite{nsvvz:78a,
nsvvz:78b}, whereas the extraction of $f_B$ from QCD sum rules was investigated
in \cite{rry:81,ae:83,nar:87,rei:88,dp:91,bbbd:92,neu:92,kr:98,nar:98}. The
first determination of $f_B$ \cite{rry:81} dates back already twenty years.
Nevertheless, due to recent theoretical progress, we find it legitimate to
reconsider this problem. Very recently, the perturbative three-loop order
$\alpha_s^2$ correction to the correlation function with one heavy and one
massless quark has been calculated \cite{cs:01,cs:01a} for the first time. It
turns out that in the pole mass scheme, which was used for most previous
analyses, due to renormalon problems \cite{bb:94}, the perturbative expansion
is far from converging. However, taking the quark mass in the $\MSb$ scheme
\cite{bbdm:78}, a very reasonable behaviour of the higher orders is obtained
and a reliable determination of $f_B$ becomes feasible.\footnote{After
completion of this work, we became aware of an independent analysis on the
same subject \cite{ps:01}, where also the new order $\alpha_s^2$ corrections
are included, however employing the framework of HQET.}

The starting point for the sum rule analysis is the two-point function
$\Psi(p^2)$ of two hadronic currents
\begin{equation}
\Psi(p^2) \; \equiv \; i \int \! dx \, e^{ipx} \,
\big<\Omega\vert \, T\{\,j_5(x)\,j_5(0)^\dagger\}\vert\Omega\big>\,,
\label{eq:1.1}
\end{equation}
where $\Omega$ denotes the physical vacuum and $j_{5}(x)$ will be the
divergence of the axialvector current,
\begin{equation}
\label{j5}
j_{5}(x) \; = \; (M+m):\!\bar q(x)\,i\gf\,Q(x)\!: \,,
\end{equation}
with $M$ and $m$ being the masses of $Q(x)$ and $q(x)$. In the following,
$Q(x)$ denotes the heavy quark which later will be specified to be the
bottom quark, whereas $q(x)$ can be one of the light quarks up, down or
strange. Note that the current $j_{5}(x)$ is a renormalisation invariant
operator. In the case of $(\bar ub)$ the corresponding matrix element is
given by
\begin{equation}
\label{fB}
(m_b+m_u)\langle 0|(\bar u i\gf b)(0)|B\rangle \; = \; f_B m_B^2 \,,
\end{equation}
where $m_B$ is the $B$-meson mass.

Up to a subtraction polynomial which depends on the large $p^2$ behaviour,
$\Psi(p^2)$ satisfies a dispersion relation (for the precise conditions see
\cite{bog:58}):
\begin{equation}
\label{disrel}
\Psi(p^2) \; = \;
\int\limits_0^\infty \frac{\rho(s)}{(s-p^2-i0)}\,ds + \mbox{subtractions} \,,
\end{equation}
where $\rho(s)$ is defined to be the spectral function $\rho(s)\equiv\IM\,
\Psi(s+i0)/\pi$. To suppress contributions in the dispersion integral coming
from higher excited states, it is further convenient to apply a Borel (inverse
Laplace) transformation to eq.~\eqn{disrel} which leads to\footnote{All
relevant formulae for the Borel transformation are collected in Appendix~A.}
\begin{equation}
\label{borelsr}
u\,{\cal B}_u\,\Psi(p^2) \; \equiv \; u\,\wh\Psi(u) \; = \;
\int\limits_0^\infty e^{-s/u}\rho(s)\,ds \,.
\end{equation}
${\cal B}_u$ is the Borel operator and the subtraction polynomial has been
removed by the Borel transformation. As we shall discuss in detail below,
the left-hand side of this equation is calculable in renormalisation group
improved perturbation theory in the framework of the operator product
expansion, if the Borel parameter $u$ can be chosen sufficiently large.

Under the {\em crucial} assumption of quark-hadron duality, the right-hand
side of eq.~\eqn{borelsr} can be evaluated in a hadron-based picture, still
maintaining the equality, and thereby relating hadronic quantities like masses
and decay widths to the fundamental Standard Model parameters. Generally,
however, from experiments the phenomenological spectral function $\rho_{ph}(s)$
is only known from threshold up to some energy $s_0$. Above this value, we
shall use the theoretical expression $\rho_{th}(s)$ also for the right-hand
side. In the case of the $B$-mesons, we approximate the phenomenological
spectral function by the pole of the lowest lying hadronic state plus the
theoretical spectral function above the threshold $s_0$,
\begin{equation}
\label{rhoph}
\rho_{ph}(s) \;=\; m_B^4 f_B^2\,\delta(s-m_B^2) + \theta(s-s_0)\rho_{th}(s) \,.
\end{equation}
This is legitimate if $s_0$ is large enough so that perturbation theory is
applicable. The central equation of our sum-rule analysis for $f_B$ then takes
the form:
\begin{equation}
\label{fbsr}
m_B^4 f_B^2 \;=\; \int\limits_0^{s_0} e^{(m_B^2-s)/u}\rho_{th}(s)\,ds \,.
\end{equation}

Besides the sum rule of eq.~\eqn{fbsr}, in our numerical analysis we shall also
utilise a second sum rule which arises from differentiating eq.~\eqn{borelsr}
with respect to $1/u$:
\begin{equation}
\label{mbsr}
-\,\frac{d}{d(1/u)}\left[u\,\wh\Psi(u)\right] \; = \; \int\limits_0^\infty s\,
e^{-s/u}\rho(s)\,ds \;=\; m_B^6 f_B^2\,e^{-m_B^2/u} + \int\limits_{s_0}^\infty
s\,e^{-s/u}\rho_{th}(s)\,ds \,.
\end{equation}
Taking the ratio of the sum rules of eqs.~\eqn{mbsr} and \eqn{fbsr}, the decay
constant drops out, and, as far as the phenomenological side is concerned, we
end up with a sum rule which only depends on the heavy meson mass $m_B$. In our
numerical analysis, this additional sum rule will be used to fix the continuum
threshold $s_0$ from the experimental value of $m_B$. The resulting $s_0$ is
then used in the $f_B$ sum rule of eq.~\eqn{fbsr}.

In section~2, we give the expressions for the perturbative pseudoscalar
spectral function up to the next-next-to-leading order in the strong
coupling, and in section~3, the non-perturbative condensate contributions
are summarised. Section~4 contains our numerical analysis of the sum rules.
Finally, in section 5, we compare our results to previous determinations
of $f_B$ in the literature and we present an estimate of the hadronic
$B$-parameter in the $B$-meson system $B_B$.

\newsection{Perturbative spectral function}

In perturbation theory, the pseudoscalar spectral function has an expansion
in powers of the strong coupling constant,
\begin{equation}
\label{rhopt}
\rho(s) \; = \; \rho^{(0)}(s) + \rho^{(1)}(s)\,a(\mu_a) +
\rho^{(2)}(s)\,a(\mu_a)^2 + \ldots \,,
\end{equation}
with $a\equiv \alpha_s/\pi$. The leading order term $\rho^{(0)}(s)$ results
from a calculation of the bare quark-antiquark loop and is given by
\begin{equation}
\label{rho0}
\rho^{(0)}(s) \; = \; \frac{N_c}{8\pi^2}\,(M+m)^2\,s\,\biggl(1-\frac{M^2}{s}
\biggr)^2 \,.
\end{equation}
For the moment, we have only kept the small quark mass $m$ in the global
factor $(M+m)^2$ and have set it to zero in the subleading contributions.
Higher order corrections in $m$ up to order $m^4$ will be discussed further
below.

Our expressions for the spectral function always implicitly contain a
$\theta$-function which specifies the starting point of the cut in the
correlator $\Psi(s)$. Although generally, we prefer to utilise the $\MSb$
mass, in order to have a scale independent starting point of the cut, in this
case we chose the pole mass $M_{{\rm pole}}$. Modulo higher order corrections,
it is always possible to rewrite the mass in the logarithms which produce the
cut in terms of the pole mass such that the $\theta$-function takes the form
$\theta(s-M_{{\rm pole}}^2)$.

The order $\alpha_s$ correction for the two point function $\Psi(s)$ was for
the first time correctly calculated in ref.~\cite{bro:81}, keeping complete
analytical dependencies in both masses $M$ and $m$. Further details on the
calculation can also be found in ref.~\cite{gen:90}. From these results it
is a simple matter to obtain the corresponding imaginary part:
\begin{eqnarray}
\label{rho1}
\rho^{(1)}(s) &\!=\!& \frac{N_c}{16\pi^2}\,C_F\, (M+m)^2\,s\,(1-x)
\Biggl\{(1-x)\biggl[\,4L_2(x)+2\ln x\ln(1-x) \\
\smvs
&&\hspace{-6mm} -\,(5-2x)\ln(1-x)\,\biggr] + (1-2x)(3-x)\ln x + 3(1-3x)\ln
\frac{\mu_m^2}{M^2} + \frac{1}{2}(17-33x)\,\Biggr\} \nn \,,
\end{eqnarray}
where $x\equiv M^2/s$ and $L_2(x)$ is the dilogarithmic function \cite{lew:81}.
The explicit form of the first order correction is sensitive to the definition
of the quark mass at the leading order. Eq.~\eqn{rho1} corresponds to running
quark masses in the $\MSb$ scheme, $M(\mu_m)$ and $m(\mu_m)$, evaluated a the
scale $\mu_m$.

The term proportional to $\ln\mu_m^2/M^2$ cancels the scale
dependence of the mass at the leading order, reflecting the fact that $\rho(s)$
is a physical quantity, i.e., independent of the renormalisation scale and
scheme. Transforming the quark mass into the pole mass scheme,\footnote{
Explicit expressions for the relation between pole and $\MSb$ mass are
collected in Appendix~B.} the resulting expression becomes scale independent
and of course agrees with eq.~(4) of \cite{ae:83}. As shall be discussed in
more detail below, however, the perturbative corrections to $f_B$ in the pole
mass scheme turn out to be rather large and we refrain from performing a
numerical analysis of the sum rule in this scheme. Therefore, our expressions
for the spectral function will only be presented in the $\MSb$ scheme.

The three-loop, order $\alpha_s^2$ correction $\rho^{(2)}(s)$ has only been
calculated very recently by Chetyrkin and Steinhauser \cite{cs:01,cs:01a} for
the case of one heavy and one massless quark. A completely analytical
computation of the second order two-point function is currently not feasible.
However, one can construct a semi-numerical approximation for $\rho^{(2)}(s)$ by
using Pad{\'e} approximations together with conformal mappings into a suitable
kinematical variable \cite{bft:93,cks:96}. The input used in this procedure is
the knowledge of eight moments for the correlator for large momentum $x\to 0$,
seven moments for small momentum $x\to\infty$, and partial information on the
threshold behaviour $x\to 1$. In our analysis, we have made use of the program
{\em Rvs.m} which contains the required expressions for $\rho^{(2)}(s)$ and
was kindly provided to the public by the authors of \cite{cs:01,cs:01a}.

In ref. \cite{cs:01,cs:01a}, the pseudoscalar spectral function $\rho(s)$ has
been calculated in the pole mass scheme. Thus we still have to add to
$\rho^{(2)}(s)$ the contributions which result from rewriting the pole mass
in terms of the $\MSb$ mass. The two contributions $\Delta_1\rho^{(2)}$ and
$\Delta_2\rho^{(2)}$ which arise from the leading and first order
contributions, respectively, are given by
\begin{eqnarray}
\label{del1rho2}
\Delta_1\rho^{(2)}(s) &\!=\!& \phantom{-}\,\frac{N_c}{8\pi^2}\, (M+m)^2\,s\,
\Big[\,(3-20x+21x^2)\,r_m^{(1)^2} - 2(1-x)(1-3x)\,r_m^{(2)}\,\Big] \,, \\
\vbox{\vskip 9mm}
\label{del2rho2}
\Delta_2\rho^{(2)}(s) &\!=\!& -\,\frac{N_c}{8\pi^2}\,C_F\, (M+m)^2\,s\,
r_m^{(1)}\biggl\{(1-x)(1-3x)\Big[\,4L_2(x)+2\ln x\ln(1-x)\,\Big] \nn \\
\vbox{\vskip 6mm}
&&\hspace{-0mm} -\,(1-x)(7-21x+8x^2)\ln(1-x) + (3-22x+29x^2-8x^3)\ln x \\
\vbox{\vskip 9mm}
&&\hspace{-0mm} +\,\frac{1}{2}(1-x)(15-31x)\,\biggr\} \nn \,.
\end{eqnarray}
Explicit expressions for the coefficients $r_m^{(1)}$ and $r_m^{(2)}$ can be
found in Appendix~B. Furthermore, in ref.~\cite{cs:01,cs:01a} the
renormalisation scale of the coupling $\mu_a$ was set to $M_{{\rm pole}}$.
Since in our numerical analysis we plan to vary the scale $\mu_a$ independently
from $\mu_m$, the contribution which results from reexpressing $a(M)$ in terms
of $a(\mu_a)$ in the two-loop part needs to be included as well.

Close to threshold, in the pole mass scheme, the pseudoscalar spectral function
behaves as $v^2 (\alpha_s\!\ln v)^k$ where $v\equiv (1-x)/(1+x)$ at any order
$k$ in perturbation theory. This behaviour, however, does not persist in the
$\MSb$ scheme, where for each order, an additional factor of $1/v$ is obtained,
such that the order $\alpha_s^2$ correction goes like a constant for $v\to 0$.
Nevertheless, as we will see in more detail below, numerically the corrections
for the integrated spectral function show a much better convergence than in
the pole mass scheme.

Let us now come to a discussion of the corrections in the small mass $m$.
At the leading order in the strong coupling and up to order $m^4$, they can,
for example, be found in ref.~\cite{jm:93}:
\begin{equation}
\label{rho0m}
\rho^{(0)}_m(s) \;=\; \frac{N_c}{8\pi^2}\,(M+m)^2\,\Biggl\{ 2(1-x)M m -
2 m^2 - 2\,\frac{(1+x)}{(1-x)}\frac{M m^3}{s} + \frac{(1-2x-x^2)}{(1-x)^2}
\frac{m^4}{s} \,\Biggr\} \,.
\end{equation}
The somewhat bulky expressions for the first order $\alpha_s$ correction can
be obtained by expanding the results of \cite{bro:81,gen:90} in terms of $m$
and have been relegated to Appendix~C. Numerically, the size of the order
$\alpha_s$ corrections increases with increasing order in the expansion in
$m$. However, even for the case of $B_s$ the mass corrections in $m_s$ become
negligible before the perturbative expansion for these corrections breaks down.

In the process of performing the expansion of the results of \cite{bro:81,
gen:90} in terms of $m$, it is found that starting with order $m^3$
logarithmic terms of the form $\ln m$ appear in the expansion. They are of
infrared origin, and in the framework of the operator product expansion it
should be possible to absorb them by a suitable definition into the higher
dimensional operator corrections, the vacuum condensates. If the operator
product expansion is performed in terms of non-normal ordered, minimally
subtracted condensates rather then the more commonly used normal ordered ones,
the mass logarithms indeed disappear \cite{sc:88,jm:93,cdps:95}.

\newsection{Condensate contributions}

In the following, we summarise the contributions to the two-point function
coming from higher dimensional operators which arise in the framework of the
operator product expansion and parametrise the appearance of non-perturbative
physics, if the energy approaches the confinement region. Here, we decided to
present directly the integrated quantity $u\wh\Psi(u)$ because the spectral
functions corresponding to the condensates contain $\delta$-distribution 
contributions.

The leading order expression for the dimension-three quark condensate is known
since the first works on the pseudoscalar heavy-light system \cite{ae:83}:
\begin{equation}
\label{qq0}
u\wh\Psi^{(0)}_{\bar qq}(u) \;=\; -\,(M+m)^2 M\qq\,
e^{-M^2/u}\,\Biggl[\, 1 - \biggl(1+\frac{M^2}{u}\biggr)\frac{m}{2M} +
\frac{M^2 m^2}{2u^2} \,\Biggr] \,.
\end{equation}
To estimate higher order mass corrections in our numerical analysis, we
have included the corresponding expansion up to order $m^2$ \cite{jm:93}.
From the mass logarithms of the perturbative order $\alpha_s$ and $m^3$
correction, it is a straightforward matter to also deduce the first order
correction to the quark condensate since the mass logarithms must cancel
once the quark condensate is expressed in terms of the non-normal ordered
condensate \cite{sc:88,jm:93,cdps:95}. We were not able to find the following
result in the literature and assume that it is new:
\begin{equation}
\label{qq1}
u\wh\Psi^{(1)}_{\bar qq}(u) \;=\; \frac{3}{2}\,C_F\,a\,(M+m)^2 M\qq\,
\Biggl\{\Gamma\biggl(0,\frac{M^2}{u}\biggr) - \biggl[\, 1 + \biggl( 1 -
\frac{M^2}{u}\biggr)\biggl(\ln\frac{\mu_m^2}{M^2}+\frac{4}{3}\biggr)\,\biggr]\,
e^{-M^2/u}\,\Biggr\} \,,
\end{equation}
where $\Gamma(n,z)$ is the incomplete $\Gamma$-function. Again, the term
$\ln\mu_m^2/M^2$ cancels the scale dependence of the mass at the leading order.

The next contribution in the operator product expansion is the dimension-four
gluon condensate. Although its influence on the heavy-light sum rule turns out
to be very small, we have nevertheless included it in the analysis. The
corresponding expression for the Borel transformed correlator is given by
\begin{equation}
\label{FF0}
u\wh\Psi^{(0)}_{FF}(u) \;=\; \frac{1}{12}\,(M+m)^2\,\FF\,e^{-M^2/u} \,.
\end{equation}
In some earlier works on the pseudoscalar sum rule this contribution appears
with a wrong sign \cite{ae:83,nar:87,nar:98}, although of course this has
negligible influence on the numerical results.

The last condensate contribution that we consider in this work is the
dimension-five mixed quark gluon condensate which still has some influence
on the sum rule since it is enhanced by the heavy quark mass. Again here
the result is well known from the literature and we just cite it for the
convenience of the reader:
\begin{equation}
\label{qFq0}
u\wh\Psi^{(0)}_{\bar qFq}(u) \;=\; -\,(M+m)^2\,\frac{M\langle g_s\bar q\sigma
Fq \rangle}{2u}\,\biggl(1-\frac{M^2}{2u}\biggr)\,e^{-M^2/u} \,.
\end{equation}
We have checked explicitly that the contribution of the next-higher dimensional
operator, the four-quark condensate, is extremely small, and thus have 
neglected all higher dimensional operators. The corresponding results for the
condensate contributions to the sum rule of eq.~\eqn{mbsr} can be calculated
straightforwardly by differentiating the above expressions with respect to
$1/u$.

\newsection{Numerical analysis}

In our numerical analysis of the pseudoscalar heavy-light sum rule, we shall
mainly discuss the values of our input parameters, their errors, and the impact
of those errors on the values of $f_B$ and $f_{B_s}$. To begin, however, let
us investigate the behaviour of the perturbative expansion.

As was already mentioned above, in the pole mass scheme the first two order
$\alpha_s$ and $\alpha_s^2$ corrections to $\wh\Psi(u)$ are of similar size 
than the leading term, thus not showing any sign of convergence. For central
values of our input parameters and a typical value $u = 5\;\gev^2$, the first
order correction amounts to 78\% and the second order to 85\% of the leading
term. To be consistent with the perturbative result for $\rho(s)$, we have used
$m_b^{{\rm pole}}=4.82\;\gev$, which results from relation \eqn{eq:b.4} up to
order $\alpha_s^2$. Because of the large corrections, we shall not pursue an
analysis in the pole mass scheme any further. On the contrary, in the $\MSb$
scheme for $\mu_m = \mu_a = m_b$ and  $u = 5\;\gev^2$, the first and second
order corrections are 11\% and 2\% of the leading term respectively, while at
$\mu_m\approx 4.5\;\gev$ the second order term vanishes entirely. Hence, in
the $\MSb$ scheme the perturbative expansion converges rather well and is
under good control.

\begin{figure}[thb]
\centerline{
\rotate[r]{
\epsfysize=14cm\epsffile{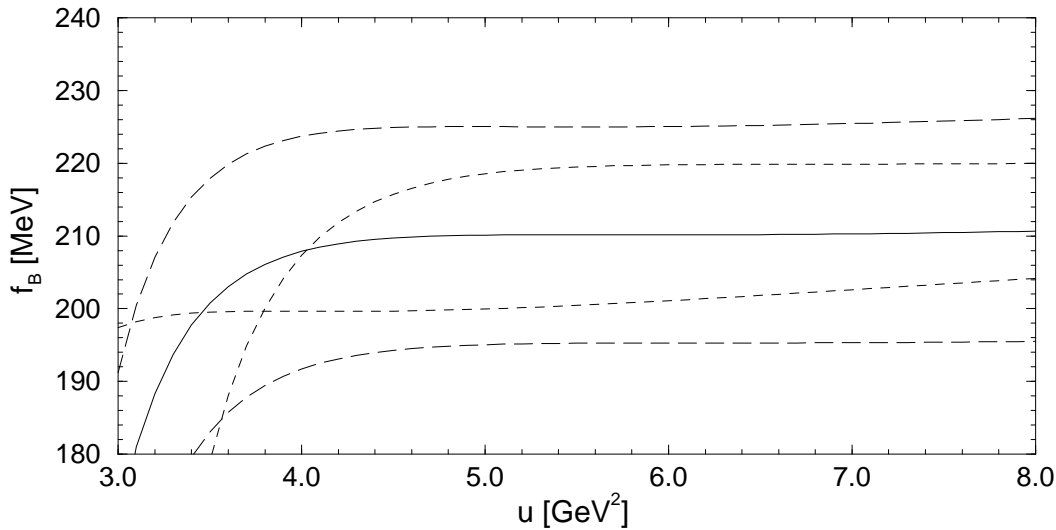} } }
\vspace{-4mm}
\caption[]{$f_B$ as a function of the Borel parameter $u$ for different sets
of input parameters. Solid line: central values of table~1; long-dashed line:
$m_b(m_b) = 4.16\;\gev$ (upper line), $m_b(m_b) = 4.26\;\gev$ (lower line);
dashed line: $\mu_m = 3\;\gev$ (lower line), $\mu_m = 6\;\gev$ (upper line).
\label{fig1}}
\end{figure}

\begin{figure}[thb]
\centerline{
\rotate[r]{
\epsfysize=14cm\epsffile{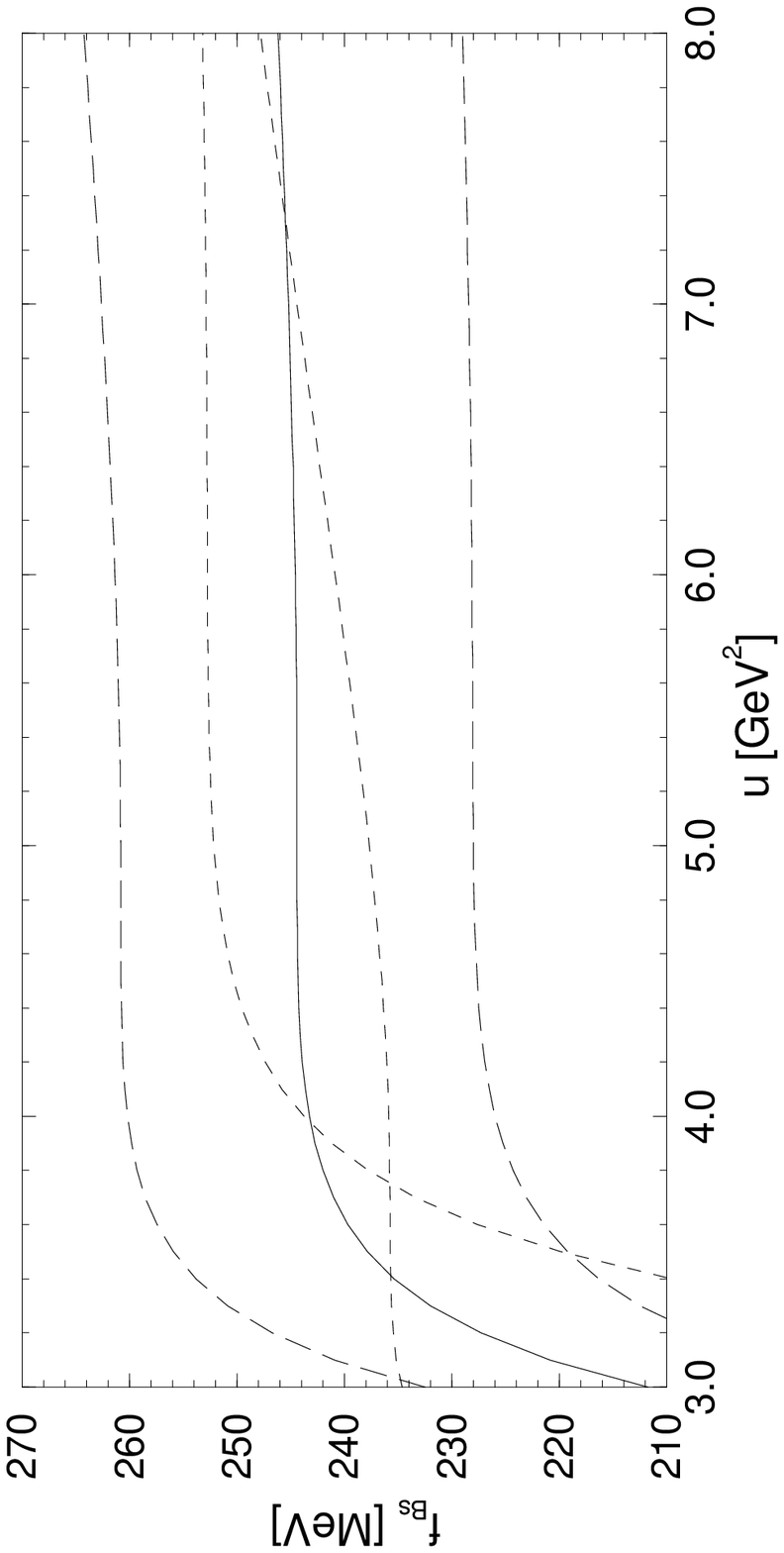} } }
\vspace{-4mm}
\caption[]{$f_{B_s}$ as a function of the Borel parameter $u$ for different sets
of input parameters. Solid line: central values of table~2; long-dashed line:
$m_b(m_b) = 4.16\;\gev$ (upper line), $m_b(m_b) = 4.26\;\gev$ (lower line);
dashed line: $\mu_m = 3\;\gev$ (lower line), $\mu_m = 6\;\gev$ (upper line).
\label{fig2}}
\end{figure}

\begin{table}[thb]
\begin{center}
\begin{tabular}{ccccc}
\hline
Parameter & Value & $s_0\;[\gev^2]$ & $u_0\;[\gev^2]$ & $\Delta f_B\;[\mev]$ \\
\hline
$m_b(m_b)$ & $4.21 \pm 0.05\;\gev$ & ${}^{33.1}_{34.2}$ & ${}^{6.1}_{5.2}$ &
$\mp 15$ \\
$\mu_m$ & $3.0-6.0\;\gev$ & ${}^{33.5}_{34.4}$ & ${}^{6.8}_{4.0}$ & $\pm10$ \\
$\mu_a$ & $3.0-6.0\;\gev$ & ${}^{34.2}_{33.1}$ & ${}^{5.1}_{6.2}$ &
${}^{+2}_{-1}$ \\
$\qq(2\;\gev)$ & $-\,(267\pm17\;\mev)^3$ & ${}^{33.9}_{33.3}$ & ${}^{5.7}_{5.5}$
 & $\pm 6$ \\
${\cal O}(\alpha_s^2)$ & ${}^{2\times{\cal O}(\alpha_s^2)}_{{\rm no}\;
{\cal O}(\alpha_s^2)}$ & -- & -- & $\pm 2$ \\
$\alpha_s(M_Z)$ & $0.1185 \pm 0.020$ & -- & -- & $\pm 1$ \\
$\FF$ & $0.024\pm 0.012\;\gev^4$ & -- & -- & $\pm 1$ \\
$m_0^2$ & $0.8 \pm 0.2\;\gev^2$ & -- & -- & $\mp 1$ \\
\hline
\end{tabular}
\end{center}
\caption{Values for all input parameters, continuum thresholds $s_0$, points
of maximal stability $u_0$, and corresponding uncertainties for $f_B$.
\label{tab:1}}
\end{table}

\begin{table}[thb]
\begin{center}
\begin{tabular}{ccccc}
\hline
Parameter & Value & $s_0\;[\gev^2]$ & $u_0\;[\gev^2]$ & $\Delta f_{B_s}\;
[\mev]$ \\
\hline
$m_b(m_b)$ & $4.21 \pm 0.05\;\gev$ & ${}^{34.8}_{36.4}$ & ${}^{5.4}_{4.8}$ &
$\mp 16$ \\
$\mu_m$ & $3.0-6.0\;\gev$ & ${}^{35.2}_{37.2}$ & ${}^{6.2}_{3.6}$ &
 ${}^{+8}_{-9}$ \\
$\mu_a$ & $3.0-6.0\;\gev$ & ${}^{36.2}_{34.9}$ & ${}^{4.7}_{5.5}$ &
$+1$ \\
$\langle\bar ss\rangle/\qq$ & $0.8 \pm 0.3$ & ${}^{35.9}_{35.2}$ &
${}^{5.3}_{4.7}$ & $\pm 8$ \\
$\qq(2\;\gev)$ & $-\,(267\pm17\;\mev)^3$ & ${}^{35.7}_{35.3}$ & ${}^{5.2}_{4.9}$
 & ${}^{+5}_{-4}$ \\
$m_s(2\;\gev)$ & $100 \pm 15\;\mev$ & -- & -- & $\pm 2$ \\
${\cal O}(\alpha_s^2)$ & ${}^{2\times{\cal O}(\alpha_s^2)}_{{\rm no}\;
{\cal O}(\alpha_s^2)}$ & -- & -- & $\pm 3$ \\
$\alpha_s(M_Z)$ & $0.1185 \pm 0.020$ & -- & -- & $\pm 1$ \\
$\FF$ & $0.024\pm 0.012\;\gev^4$ & -- & -- & $\pm 1$ \\
$m_0^2$ & $0.8 \pm 0.2\;\gev^2$ & -- & -- & $\mp 1$ \\
\hline
\end{tabular}
\end{center}
\caption{Values for all input parameters, continuum thresholds $s_0$, points
of maximal stability $u_0$, and corresponding uncertainties for $f_{B_s}$.
\label{tab:2}}
\end{table}

In figs.~1 and 2, as the solid lines we display the leptonic decay constants
$f_B$ and $f_{B_s}$ for central values of all input parameters which have been
collected in tables~1 and 2, as a function of the Borel variable $u$. For
$u\lesssim 4\;\gev^2$ the power corrections become comparable to the
perturbative term, whereas for $u\gtrsim 6\;\gev^2$ the continuum contribution
gets as important as the phenomenological part. Thus a reliable sum rule
analysis should be possible in the range roughly given by $4\;\gev^2 \lesssim
u \lesssim 6\;\gev^2$. In this region we extract our central results
$f_B=210\;\mev$ and $f_{B_s}=244\;\mev$.

\begin{figure}[thb]
\centerline{
\rotate[r]{
\epsfysize=14cm\epsffile{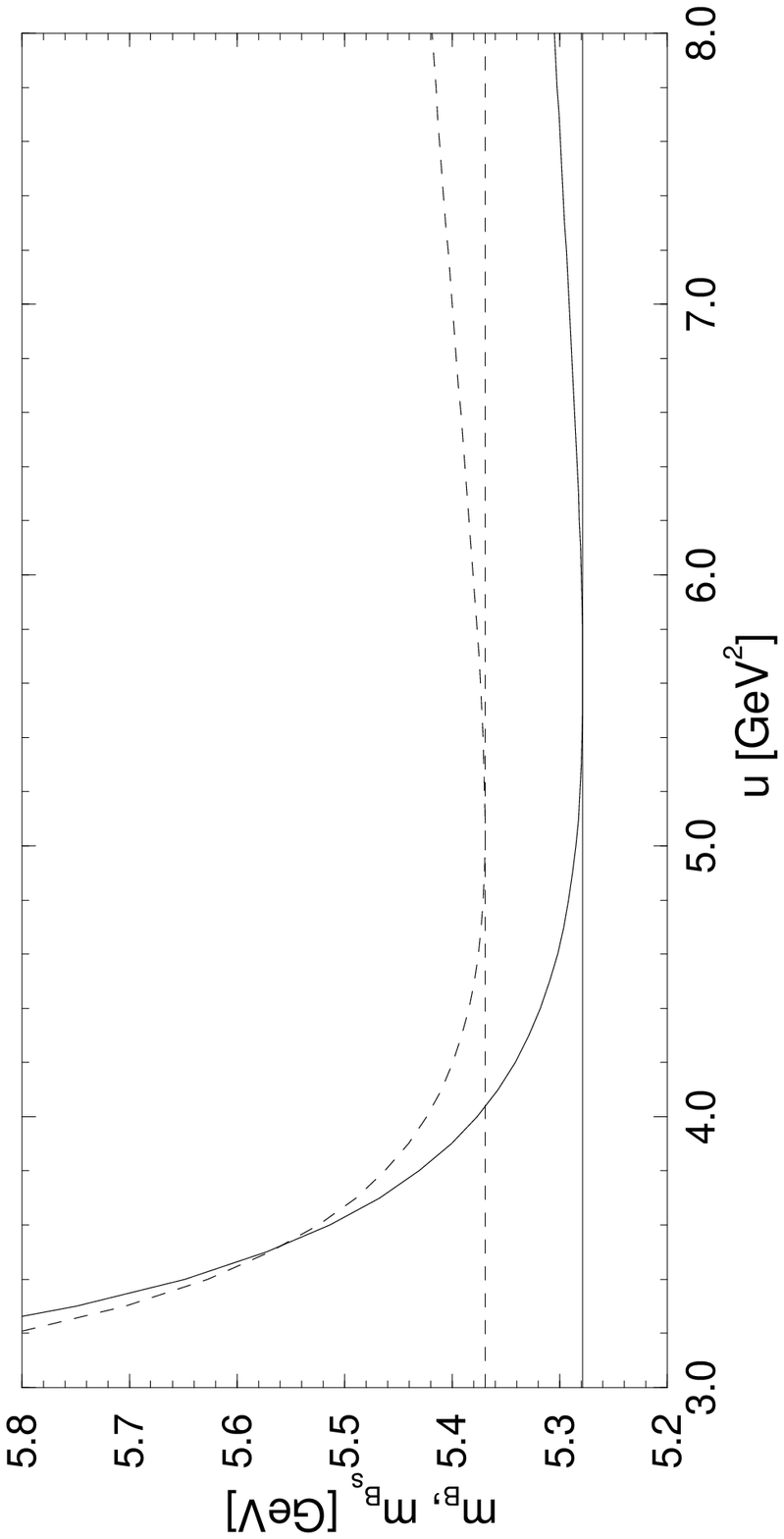} } }
\vspace{-4mm}
\caption[]{$m_B$ (solid line) and $m_{B_s}$ (dashed line) as a function of the
Borel parameter $u$ for central input parameters. The horizontal lines indicate
the corresponding experimental values for these quantities.
\label{fig3}}
\end{figure}

As an additional input parameter the continuum threshold $s_0$ is required.
This parameter can be determined from the ratio of the sum rules of eqs.
\eqn{mbsr} and \eqn{fbsr}, which only depends on the heavy meson mass. To this
end, for a certain set of input parameters, $s_0$ is tuned such as to reproduce
the Particle Data Group values for $m_B$ and $m_{B_s}$ \cite{pdg:00} in the
stability region (a minimum in this case) of the ratio of sum rules. In
tables~1 and 2, we also present the resulting values for $s_0$ and the
corresponding location $u_0$ of the minimum of the $m_B$ sum rule. For central
values of all input parameters, we obtain $s_0=33.6\;\gev^2$ and $u_0=5.6\;
\gev^2$ for the $B$-meson, as well as $s_0=35.5\;\gev^2$ and $u_0=5.1\;\gev^2$
for the $B_s$-meson. In fig.~\ref{fig3}, we show the resulting $m_B$ and
$m_{B_s}$ as a function of $u$ for central input parameters. As can be seen
from this figure, in the stability region, the sum rule reproduces the physical
heavy meson masses which are indicated as horizontal lines. Our results for
$f_B$ and $f_{B_s}$ are then extracted at $u_0$, around which also the sum
rules for the decay constants are most stable and display an inflection point.

The dominant source of uncertainty for the decay constants is the error on the
bottom quark mass $m_b$. For this value we have taken an average over recent
determinations \cite{jp:97,kpp:98,my:99,pp:99,bs:99,hoa:00,gim:00,pin:01,ks:01}
which results in $m_b(m_b) = 4.21 \pm 0.05\;\gev$. The error on $m_b$ has been
chosen such that all individual results are included within one standard
deviation.  The corresponding variations of $f_B$ and $f_{B_s}$ are displayed
as the long-dashed lines in figs.~1 and 2, where the upper line corresponds to
a lower value of $m_b$ and the lower line to a larger $m_b$. The impact of the
variation of $m_b$ on the error of $f_B$ and $f_{B_s}$ has been quantified in
tables~1 and 2.

Another important source of uncertainty is the renormalisation scale $\mu_m$.
We have decided to vary $\mu_m$ in the range $3$ -- $6\;\gev$, with a central
value $\mu_m=m_b$. If $\mu_m$ is smaller than about $3\;\gev$, the perturbative
corrections become too large and the expansion unreliable. As the dashed lines
in figs.~1 and 2, we then show the corresponding results for $\mu_m=3\;\gev$
(lower line) and $\mu_m=6\;\gev$ (upper line). The uncertainties for $f_B$ and
$f_{B_s}$ which result from $\mu_m$ are again listed in tables~1 and 2. To
indicate the influence of even lower scales, let us briefly discuss the case
$\mu_m=2.5\;\gev$. Here, we find $s_0=38.4\;\gev^2$ being rather large, as well
as $u_0=2.9\;\gev^2$ which is very small. At such a low $u_0$, the perturbative
and operator product expansions are not very reliable. Nevertheless, the value
for $f_B$ extracted at $u_0$ turns out surprisingly close to our central result,
such that the error estimate of table~1 is more conservative. The variation of
$\mu_a$, on the other hand, only has a minor impact on the error of $f_B$ and
$f_{B_s}$ and is also given in tables~1 and 2.

The present uncertainties in the remaining QCD parameters $\alpha_s$, the
strange quark mass $m_s$ and the condensate parameters have much less
influence on the errors of $f_B$ and $f_{B_s}$. Thus let us be more brief with
the discussion of these quantities. The current value of $\alpha_s(M_Z)$ by the
Particle Data Group, $\alpha_s(M_Z)=0.1185 \pm 0.020$ \cite{pdg:00}, has been
used, whereas our choice for the strange mass $m_s(2\;\gev)=100\pm 15 \;\mev$
is obtained from two very recent analyses of scalar and pseudoscalar QCD sum
rules \cite{mk:01, jop:01}. The resulting $m_s$ is compatible to the
determination from hadronic $\tau$-decays, as well as lattice QCD results
\cite{gm:01,che:01,kan:01}. Besides the variation of $\alpha_s(M_Z)$, in order
to estimate the influence of higher order corrections, we have either removed
or doubled the known ${\cal O}(\alpha_s^2)$ correction. The resulting
uncertainty for the decay constants, however, turns out to be small.

Our value for the quark condensate has been extracted
from the Gell-Mann-Oakes-Renner relation \cite{gmor:68} with current values for
the up- and down-quark masses \cite{jop:01}. The ratio $\langle\bar ss\rangle/
\qq$ has been chosen such as to include results from refs.~\cite{djn:89,nar:89,
ls:91,jm:95,drs:01}.\footnote{We have not taken into account the very recent
result $\langle\bar ss\rangle/\qq=1.7$ \cite{abt:01}, obtained in the framework
of chiral perturbation theory ($\chi$PT), which would lead to $f_{B_s}=270\;
\mev$. In $\chi$PT, the value of the quark condensate depends on the subtraction
procedure employed, and it is not clear how these results relate to $\qq$ in
the $\MSb$ scheme. The large value obtained in \cite{abt:01} can almost be
excluded on the basis of our $f_B$, together with independent lattice results
for the ratio $f_{B_s}/f_B$ (see below).} The mixed quark-gluon condensate is
parametrised by $\langle g_s\bar q\sigma Fq \rangle = m_0^2 \qq$ with $m_0^2$
being determined in ref.~\cite{op:88}, and finally, for the gluon condensate
we take a generous range which includes previous values found in the literature.
All uncertainties for $f_B$ and $f_{B_s}$ resulting from these parameters are
also listed in tables~1 and 2. Where entries for $s_0$ and $u_0$ are missing,
we have used the values corresponding to central input parameters.

Adding all errors for the various input parameters in quadrature, our
final results for the $B$ and $B_s$ meson leptonic decay constants are:
\begin{equation}
\label{fbfbs}
f_B \;=\; 210 \pm 19 \; \mev
\quad\qquad \mbox{and} \quad\qquad
f_{B_s} \;=\; 244 \pm 21 \; \mev \,.
\end{equation}
In the next section, we shall compare these values with previous QCD sum rule
and lattice QCD determinations.

\newsection{Conclusions}

The only truly non-perturbative method to compute hadronic matrix elements
is QCD on a space-time lattice and thus it is very interesting to compare our
findings to the corresponding results in lattice gauge theory. For the leptonic
heavy meson decay constants, they have been compiled in a recent review article
by Bernard \cite{ber:00}.\footnote{See also ref. \cite{sac:01}.} Taking into
account dynamical sea quark effects and estimating the corresponding
uncertainties, his world averages read:
\begin{equation}
\label{fblat}
f_B \;=\; 200 \pm 30 \; \mev
\qquad \mbox{and} \qquad
\frac{f_{B_s}}{f_B} \;=\; 1.16 \pm 0.04 \,.
\end{equation}
The lattice value for $f_B$ is in good agreement with our result of
eq.~\eqn{fbfbs}, and also our ration $f_{B_s}/f_B=1.16$ turns out to be
perfectly consistent with \eqn{fblat}. Nevertheless, due to sizable
discretisation errors on the lattice, in our opinion, at present the QCD
sum rule determination of the decay constants is more precise.

We now come to a comparison with recent QCD sum rule results for $f_B$
and $f_{B_s}$. The status of sum rule calculations of $f_B$ in the pole mass
scheme has been summarised in the review article \cite{kr:98} with the
result $f_B = 180\pm 30\;\mev$. Although roughly 15\% lower, within the errors
this result is compatible with our value \eqn{fbfbs}. However, due to the
large perturbative corrections in the pole mass scheme, and the strong
dependence on the bottom quark mass which in \cite{kr:98} was taken to be
$m_b^{{\rm pole}}=4.7\pm 0.1\;\gev$, the theoretical error is not controlled
reliably. Let us remark that the order $\alpha_s^3$ correction in the relation
between $\MSb$ mass and pole mass alone gives a shift of $m_b^{{\rm pole}}$
by roughly $200\;\mev$ \cite{cs:99,cs:00,mr:00}. Typical results for $f_{B_s}$
turn out to be about $35\;\mev$ higher than $f_B$ \cite{kr:98}, so that the
difference between $f_{B_s}$ and $f_B$ is in agreement to our result. Our
result for $f_B$ is also completely compatible with the very recent analysis
of ref.~\cite{ps:01}, which was performed in the framework of HQET and resulted
in $f_B=206\pm 20\;\mev$, suffering however from the problems of the pole mass
discussed above.

After submission of our work to the e-print archive, an independent analysis of
the heavy-light meson sum rules by Narison \cite{nar:01} was published, which
also employs the heavy quark mass in the $\MSb$ scheme. For the convenience
of the reader, even though ref.~\cite{nar:01} appeared later, we have been
asked by our referee to nevertheless comment on this analysis. The main
difference to our analysis lies in the fact that in ref. \cite{nar:01} the
bottom quark mass is extracted from the sum rule for $m_B$, with the result
$m_b(m_b)=4.05\pm 0.06\;\gev$. We have checked that for this value of $m_b$
one needs $s_0=37.5\;\gev^2$ to reproduce $m_B$, and finds a stability region
around $u_0=4.3\;\gev^2$. Inserting these parameters into the $f_B$ sum rule,
we obtain $f_B=270\;\mev$, in conflict to our result \eqn{fbfbs}. We are able
to reproduce the value quoted by Narison, $f_B=205\;\mev$, at $u=2.7\;\gev^2$,
which roughly corresponds to his preferred $\tau\equiv1/u$ value. Around this
$u$, however, the $f_B$ sum rule is unstable, casting doubts on the procedure
of also demanding stability in the continuum threshold $s_0$, besides the
$u$-stability. Furthermore, the rather low value of $m_b$ compared to our world
average presented above, as well as the very high value of $s_0$, indicate that
the pseudoscalar sum rule is not a good place to determine $m_b$. On the other
hand, our investigation demonstrates that perfectly compatible results are
obtained with a more standard value for $m_b$. The ratio $f_{B_s}/ f_B$ has
been calculated by the same author in \cite{nar:94} with the result $f_{B_s}/
f_B = 1.16\pm 0.05$, in agreement to our findings.

The heavy-meson decay constant $f_B$ plays an important role in the mixing of
neutral $B^0$ and $\bar B^0$ mesons. The relevant hadronic matrix element
can be expressed as \cite{bjw:90}
\begin{equation}
\label{bbmix}
\langle\bar B^0|\wh Q_{\Delta B=2}|B^0\rangle \;=\;
\frac{8}{3}\,B_B f_B^2\,m_B^2 \,,
\end{equation}
where $\wh Q_{\Delta B=2}$ is the scale invariant four-quark operator which
mediates $B^0$-$\bar B^0$ mixing and $B_B$ is the corresponding scale
invariant $B$-parameter which parametrises the deviation of the matrix
element from the factorisation approximation. In the factorisation
approximation, by definition we would have $B_B=1$. The combination
$\sqrt{B_B}f_B$ can be extracted from an analysis of experimental data on
$B^0$-$\bar B^0$ mixing together with additional inputs which determine the
matrix elements of the quark mixing or Cabibbo-Kobayashi-Maskawa matrix. A
very recent analysis then yields $\sqrt{B_B}f_B = 236\pm 35\;\mev$
\cite{hlll:01,hoe:01}. Taking together our result for $f_B$ and the quoted
value for $\sqrt{B_B}f_B$, we are in a position to give an estimate of the
scale invariant $B$-parameter $B_B$, which reads
\begin{equation}
\label{BB}
B_B \;=\; 1.26 \pm 0.45 \,.
\end{equation}
For simplicity we have assumed Gaussian errors in both input quantities. The
result again is in very good agreement to corresponding determinations of
$B_B$ on the lattice which gave $B_B = 1.30\pm 0.12\pm 0.13$ \cite{ber:00},
although our error in this case is bigger.

To conclude, in this work we have presented a QCD sum rule determination of
the leptonic heavy-meson decay constants $f_B$ and $f_{B_s}$. Due to large
perturbative higher order corrections, an analysis in terms of the bottom
quark pole mass appeared unreliable. On the contrary, employing the heavy
quark mass in the $\MSb$ scheme, up to order $\alpha_s^2$ the perturbative
expansion displays good convergence and a reliable determination of $f_B$ and
$f_{B_s}$ turned out possible. Our central results have been presented in eq.
\eqn{fbfbs}, where the dominant uncertainty arose from the present error in the
bottom quark mass $m_b(m_b)$. Taking into account independent information on
$\sqrt{B_B}f_B$ from $B^0$-$\bar B^0$ mixing, we were also in a position to
give an estimate on the $B$-meson $B$-parameter $B_B$ in eq.~\eqn{BB}. All our
results are in very good agreement to lattice QCD determinations of the same
quantities. Further improvements of our results will only be possible if the
dominant theoretical uncertainties could be reduced. This would require a more
precise value of the bottom mass, and a reduction of the renormalisation scale
dependence, requiring the next perturbative order $\alpha_s^3$ correction,
which at present seems to be out of reach.

\vskip 1cm \noindent
{\Large\bf Acknowledgements}
 
\noindent
It is a pleasure to thank H. G. Dosch for discussions. We also thank
M.~Steinhauser and A.~Penin for pointing out a sign misprint in the
original version of eq.~\eqn{del2rho2}, and M.~J. would like to thank
the Deutsche Forschungsgemeinschaft for their support.

\newpage
\appendix{\LARGE\bf Appendices}
\newsection{The Borel transform}

The Borel operator ${\cal B}_u$ is defined by ($s\equiv -p^2$)
\label{eq:a.1}
\begin{equation}
{\cal B}_u \; \equiv \; \lim_{\stackrel{s,n\to\infty}{s/n=u}}
\frac{(-s)^n}{(n-1)!}\,\frac{\partial^{\,n}}{(\partial s)^n} \,.
\end{equation}
The Borel transformation is an inverse Laplace transform \cite{wid:46}.
If we set
\begin{equation}
\label{eq:a.2}
\wh f(u) \; \equiv \; {\cal B}_u\Big[\,f(s)\,\Big] \,,
\qquad \hbox{then} \qquad
f(s) \; = \; \int\limits_0^\infty \frac{1}{u}\,\wh f(u)\,e^{-s/u}du \,.
\end{equation}
In this work we just need the following Borel transform:
\begin{equation}
\label{eq:a.3}
{\cal B}_u\biggl[\,\frac{1}{(x+s)^\alpha}\,\biggr] \; = \;
\frac{1}{u^\alpha \Gamma(\alpha)}\,e^{-x/u} \,.
\end{equation}
Cases in which logarithms appear can be treated by first evaluating the
spectral function and then calculating the dispersion integral of
eq.~\eqn{borelsr}.

\newsection{Renormalisation group functions}

For the definition of the renormalisation group functions we
follow the notation of Pascual and Tarrach \cite{pt:84}, except
that we define the $\beta$-function such that $\beta_1$ is positive.
The expansions of $\beta(a)$ and $\gamma(a)$ take the form:
\begin{equation}
\label{eq:b.1}
\beta(a) \; = \; -\,\beta_1 a-\beta_2 a^2-\beta_3 a^3-\ldots \,,
\quad \hbox{and} \quad
\gamma(a) \; = \; \gamma_1 a+\gamma_2 a^2+\gamma_3 a^3+\ldots \,,
\end{equation}
with
\begin{equation}
\label{eq:b.2}
\beta_1 \; = \; \frac{1}{6}\,\Big[\,11C_A-4Tn_f\,\Big] \,, \qquad
\beta_2 \; = \; \frac{1}{12}\,\Big[\,17C_A^2-10C_ATn_f-6C_FTn_f\,\Big] \,,
\end{equation}
and
\begin{equation}
\label{eq:b.3}
\gamma_1 \; = \; \frac{3}{2}\,C_F \,, \qquad
\gamma_2 \; = \; \frac{C_F}{48}\,\Big[\,97C_A+9C_F-20Tn_f\,\Big] \,.
\end{equation}

The relation between pole and running $\MSb$ mass is given by
\begin{equation}
\label{eq:b.4}
M(\mu_m) \; = \; M_{{\rm pole}}\,\Big[\,1 + a(\mu_a)\,r_m^{(1)}(\mu_m) +
a(\mu_a)^2\,r_m^{(2)}(\mu_a,\mu_m) + \ldots \,\Big] \,,
\end{equation}
where
\begin{eqnarray}
r_m^{(1)} & = & r_{m,0}^{(1)} - \gamma_1\ln\frac{\mu_m}{M(\mu_m)} \,,
\label{eq:b.5} \\
\smvs
r_m^{(2)} & = & r_{m,0}^{(2)} - \Big[\,\gamma_2+(\gamma_1-\beta_1)\,
r_{m,0}^{(1)}\,\Big]\ln\frac{\mu_m}{M(\mu_m)} + \frac{\gamma_1}{2}\,(\gamma_1-
\beta_1)\ln^2\frac{\mu_m}{M(\mu_m)} \nn \\
\smvs
& & -\,\biggl[\,\gamma_1+\beta_1\ln\frac{\mu_m}{\mu_a}\,\biggr]
r_m^{(1)} \,. \label{eq:b.6}
\end{eqnarray}
The coefficients of the logarithms can be calculated from the renormalisation
group \cite{cks:96} and the constant coefficients $r_{m,0}^{(1)}$ and
$r_{m,0}^{(2)}$ are found to be \cite{tar:81,gbgs:90}
\begin{eqnarray}
r_{m,0}^{(1)} & = & -\,C_F \,, \label{eq:b.7} \\
\smvs
r_{m,0}^{(2)} & = & C_F^2\biggl(\frac{7}{128}-\frac{15}{8}\zeta(2)-\frac{3}{4}
\zeta(3)+3\zeta(2)\ln 2\biggr)+C_FTn_f\biggl(\frac{71}{96}+\frac{1}{2}\zeta(2)
\biggr) \nn \\
\smvs
& & \hspace{-16mm}+\,C_AC_F\biggl(-\frac{1111}{384}+\frac{1}{2}\zeta(2)+
\frac{3}{8}\zeta(3)-\frac{3}{2}\zeta(2)\ln 2\biggr)+C_FT\biggl(\frac{3}{4}-
\frac{3}{2}\zeta(2)\biggr) \,. \label{eq:b.8}
\end{eqnarray}

\newsection{\boldmath Mass corrections at order $\alpha_s$\unboldmath}

Below, we present the order $\alpha_s$ mass corrections to the pseudoscalar
spectral function which arise from expanding the results by \cite{bro:81,
gen:90} up to order $m^4$, after the higher dimensional operators have been
expressed in terms of non-normal ordered condensates:

\begin{eqnarray}
\label{rho1m1}
\rho^{(1)}_m(s) &\!=\!& \phantom{-}\,\frac{N_c}{8\pi^2}\,C_F\, (M+m)^2\,
M m\,\Biggl\{(1-x)\biggl[\,4L_2(x)+2\ln x\ln(1-x) \\
\smvs
&&\hspace{-6mm} -\,2(4-x)\ln(1-x)\,\biggr] + 2(3-5x+x^2)\ln x + 3(2-3x)\ln
\frac{\mu_m^2}{M^2} + 2(7-9x)\,\Biggr\} \nn \,, \\
\vbox{\vskip 12mm}
\label{rho1m2}
\rho^{(1)}_{m^2}(s) &\!=\!& -\,\frac{N_c}{8\pi^2}\,C_F\, (M+m)^2\,m^2\,
\Biggl\{(1-x)\biggl[\,4L_2(x)+2\ln x\ln(1-x)\,\biggr] \nn \\
\smvs
&&\hspace{-6mm} -\,(2+x)(4-x)\ln(1-x) + (6+2x-x^2)\ln x + 6\ln
\frac{\mu_m^2}{M^2} + (8-3x)\,\Biggr\} \,, \\
\vbox{\vskip 12mm}
\label{rho1m3}
\rho^{(1)}_{m^3}(s) &\!=\!& -\,\frac{N_c}{8\pi^2}\,C_F\, (M+m)^2\,\frac{M m^3}
{s}\,\Biggl\{\,4L_2(x)+2\ln x\ln(1-x) + \frac{(9+8x-9x^2)}{(1-x)^2} \nn \\
\smvs
&&\hspace{-6mm} -\,2\frac{(7+7x-2x^2)}{(1-x)}\ln(1-x) + 2\frac{(6+7x-2x^2)}
{(1-x)}\ln x + 6\frac{(2-x^2)}{(1-x)^2}\ln\frac{\mu_m^2}{M^2}\,\Biggr\} \,, \\
\vbox{\vskip 12mm}
\label{rho1m4}
\rho^{(1)}_{m^4}(s) &\!=\!& \phantom{-}\,\frac{N_c}{8\pi^2}\,C_F\, (M+m)^2\,
\frac{m^4}{s}\,\Biggl\{\,2L_2(x)+\ln x\ln(1-x) \nn \\
\smvs
&&\hspace{-6mm} -\,\frac{(13-24x-27x^2+2x^3)}{2(1-x)^2}\ln(1-x) +
\frac{(12-22x-27x^2+2x^3)}{2(1-x)^2}\ln x \nn \\
\smvs
&&\hspace{-6mm} +\,3\frac{(4-12x+x^2+3x^3)}{2(1-x)^3}\ln\frac{\mu_m^2}{M^2} +
\frac{(6-64x+15x^2+11x^3)}{4(1-x)^3}\,\Biggr\} \,.
\end{eqnarray}

\newpage

\end{document}